\documentstyle[tighten,prl,epsf,floats,aps]{revtex}

\begin{document}

\twocolumn[
\hsize\textwidth\columnwidth\hsize\csname @twocolumnfalse\endcsname

\draft \preprint{\today}
\title{Scattering Rate Gap in the IR Response of HgBa$_2$Ca$_2$Cu$_3$O$_{8+\delta}$}

\author{J. J. McGuire, M. Windt, T. Startseva, T. Timusk}
\address{Department of Physics and
Astronomy, McMaster University, Hamilton, Ontario L8S 4M1, Canada}

\author{D. Colson, V. Viallet-Guillen\cite{V-G}}
\address{Physique de l'Etat Condens\'e, DRECAM/SPEC, CEA, Saclay,
91191 Gif sur Yvette, France}

\maketitle

\begin{abstract}
The $ab$-plane optical spectra of one underdoped and one nearly optimally
doped single crystal of HgBa$_2$Ca$_2$Cu$_3$O$_{8+\delta}$ were
investigated in the frequency range from 40 to 40,000~cm$^{-1}$. The
frequency dependent scattering rate was obtained by Kramers Kronig
analysis of the reflectance.  Both crystals have a scattering rate gap 
of about 1000~cm$^{-1}$ which is much larger than the 700~cm$^{-1}$ gap 
seen in optical studies of several cuprates with maximum $T_c$ 
around 93~K. There appears to be a universal scaling between scattering 
rate gap and maximum $T_c$ for the cuprate superconductors.
\end{abstract}

 ]

Although high temperature superconductivity in the copper oxides was
discovered over a decade ago, an understanding of the mechanism that gives
rise to the high transition temperature, $T_c$, is still elusive.  Most of
the research has been directed towards the one- and two-layer systems which
have a maximum $T_c$ in the 93~K range. Recently, however, high quality
single crystals of the three-layer Hg based materials with a maximum $T_c
\approx 135$~K have become available\cite{Colson94} making it possible to
examine, spectroscopically, oxide superconductors that have significantly
higher $T_c$.

Of particular interest is the magnitude of the normal state pseudogap. This
is a partial suppression of the density of low energy excitations seen well
above $T_c$ in all underdoped high temperature superconductors. A pseudogap
has been seen with a variety of techniques such as angle-resolved
photoemission (ARPES), tunneling spectroscopy, specific heat, dc resistivity,
nuclear magnetic resonance and optical spectroscopy\cite{Timusk99}.
Measurements of dc resistivity\cite{Carrington94} and NMR $^{63}$Cu
$T_1$\cite{Julien96} on the three-layer underdoped
HgBa$_2$Ca$_2$Cu$_3$O$_{8+\delta}$ (Hg-1223) show evidence of a pseudogap
with onset temperatures, $T^*$, of 320~K and 230~K respectively.

The size of the pseudogap in the one- and two-layer materials is of the order
of 9.5k$_B$T$_{c}$ at optimal doping and essentially independent of
temperature. As $T_c$ is reduced below optimal doping the pseudogap increases
in size\cite{renner98,dewilde98,loram97}.

In $ab$-plane infrared response, the pseudogap is most 
clearly seen in the frequency-dependent scattering rate which can be 
calculated from the reflectance after Kramers-Kronig analysis. Here the 
pseudogap is a suppression of scattering below a characteristic energy which 
is taken to be a measure of the "size" of the pseudogap. A
comparison\cite{Puchkov96} of infrared spectra of YBa$_2$Cu$_3$O$_x$ (Y-123),
YBa$_2$Cu$_4$O$_8$ (Y-124), Bi$_2$Sr$_2$CaCu$_2$O$_{8+\delta}$ (Bi-2212) and
Tl$_2$Sr$_2$CuO$_{6+\delta}$ (Tl-2201) showed no dependence of pseudogap size
on doping level\cite{doping} or even on which system was being measured. 
All samples showed a pseudogap of $\approx$ 700~cm$^{-1}$. A relationship 
between pseudogap size and maximum $T_c$ could not be ruled out since all of 
these materials have $T_c$ near 93~K at optimal doping.

In this work the infrared reflectance of single crystal Hg-1223 is measured
in order to study the scattering rate gap\cite{gap} in a material with a 
maximum $T_c$ significantly higher than those of the other systems that have 
been studied previously\cite{Choi97}. 

The measurements were carried out on two single crystals grown in gold foil
by a single step synthesis\cite{Colson94}.  The crystals were characterized
by X-ray diffraction and wavelength dispersive
spectrometry\cite{Bertinotti97}.  The nearly optimally doped crystal had a
$T_c$ of 130~K and dimensions 0.4 $\times$ 0.3 $\times$ 0.03~mm$^3$.  The
underdoped sample had a $T_c$ of 121~K and dimensions 0.6 $\times$ 0.5
$\times$ 0.04~mm$^3$.  The smallest dimension is the $c$-axis or [001]
direction with the [100] direction $45^{\circ}$ from the two larger edges.
In both crystals the width of the transition was about 6~K. Reflectance
measurements between 40 and 8000~cm$^{-1}$ were performed on as-grown
$ab$-plane faces with a Michelson interferometer using three different
detectors.  A grating spectrometer with three additional detectors was used
for the rest of the range up to 40,000~cm$^{-1}$ (5~eV).

\begin{figure}[t]
\leavevmode
\epsfxsize=\columnwidth
\centerline{\epsffile{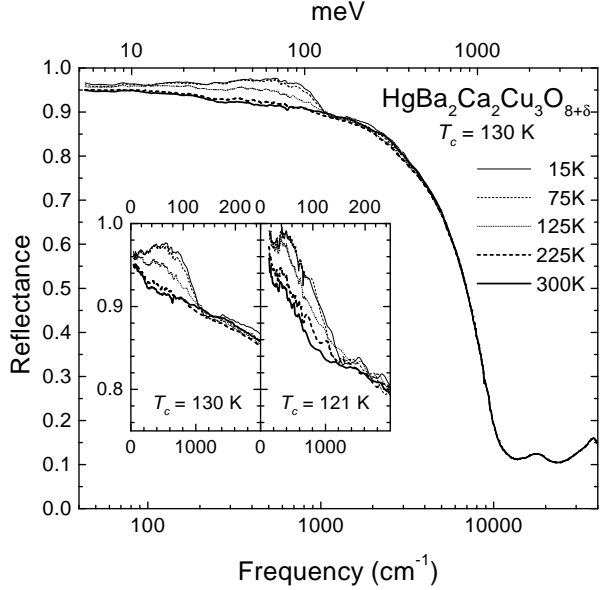}}
\vspace{0.1in}
 \caption{$ab$-plane reflectance of the optimally doped crystal of Hg-1223
with $T_c = 130$~K.  The inset compares this to the reflectance of the
underdoped sample with $T_c = 121$~K. The temperature dependence below
1000~cm$^{-1}$ is due to the scattering rate gap.} 
\label{figure1}
\end{figure}

The $ab$-plane reflectance of the optimally doped crystal is shown in Fig.\ 1.
At low temperature there is a striking temperature-dependent feature below
1000~cm$^{-1}$. In the optimally doped sample this is seen only in the three
spectra below $T_c$.  The inset of the figure shows a similar feature in the
reflectance of the underdoped sample also at 1000~cm$^{-1}$. The presence
of the feature in the 125~K data, which is now above the 121~K $T_c$, shows
that it persists in the normal state, although at a somewhat lower frequency.
A similar decrease in frequency above $T_c$ was also observed in
Bi-2212\cite{Puchkov96}. We will identify the feature at 1000~cm$^{-1}$ in 
the superconducting state of both samples as the scattering rate gap in 
analogy with other high temperature superconductors.

\begin{figure}[t]
\leavevmode
\epsfxsize=\columnwidth
\centerline{\epsffile{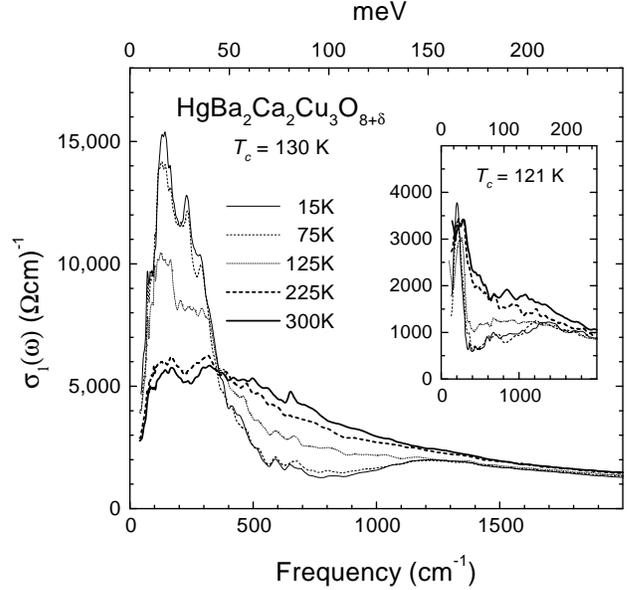}}
\vspace{0.1in}
 \caption{The real part of the optical conductivity of the optimally doped
crystal of Hg-1223. There is a marked transfer of spectral weight from
$\omega<1200$ to a peak centered at 184~cm$^{-1}$. This peak is much narrower
and weaker in the spectra shown in the inset for the underdoped sample.}
\label{figure2}
\end{figure}

The calculation of the optical conductivity and frequency-dependent scattering
rate requires extrapolation of the reflectance to all frequencies for the
Kramers-Kronig analysis.  The spectra were extrapolated to high frequencies
using the Bi-2212 data of Terasaki {\it et al.}\cite{Terasaki90prb} above
5~eV, and using power law extrapolations: $\omega^{-1}$ above 25~eV and
$\omega^{-4}$ above 124~eV. Below our lowest measured point at 43~cm$^{-1}$
we assumed a Drude conductivity in the normal state and $1-\omega^{2}$
reflectance in the superconducting state. We estimate our experimental 
uncertainty of the reflectance to be $\pm0.005$. Combined with uncertainties 
due to the extrapolations, this gives an uncertainty in our optical 
conductivity of $\pm 8\%$ above 200~cm$^{-1}$. At low frequencies the 
uncertainty rises, reaching $\pm 40\%$ at 43~cm$^{-1}$. The resolution of the 
spectra is 20~cm$^{-1}$ up to 680~cm$^{-1}$ and 30~cm$^{-1}$ up to 
8000~cm$^{-1}$.

\begin{figure}[t]
\leavevmode
\epsfxsize=\columnwidth
\centerline{\epsffile{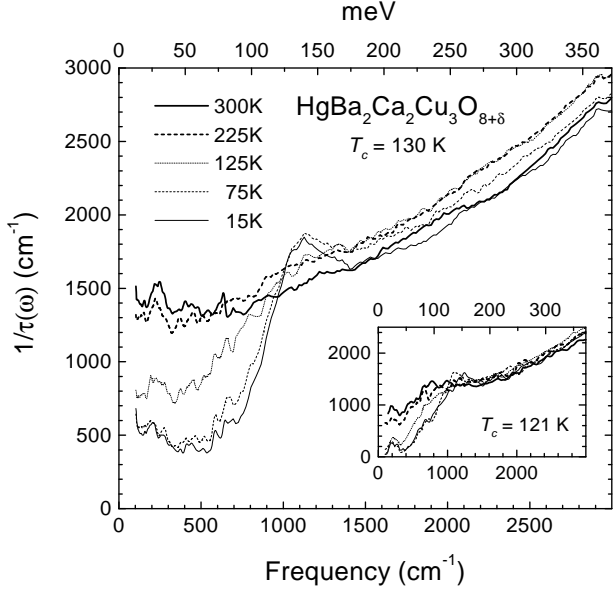}}
\vspace{0.1in}
 \caption{The frequency-dependent scattering rate of the optimally doped
crystal of Hg-1223.  The depression below 1000~cm$^{-1}$ is due to the
scattering rate gap.  The inset shows the spectra for the underdoped sample.}
\label{figure3}
\end{figure}

The real part of the optical conductivity of the optimally doped sample is
shown in Fig.\ 2.  As the temperature is lowered there is a transfer of
spectral weight from a region between 360 and 1200~cm$^{-1}$ to a peak
centered at 184~cm$^{-1}$.   There is an isosbestic point in the conductivity
spectra at 360~cm$^{-1}$ where the curves taken at different temperatures
cross suggesting that there is a transfer of spectral weight from one
component to another without an attendant change in the lineshape of either
component. In this case one component is a Drude-like continuum, and the
other is the peak at 184~cm$^{-1}$. Such a peak is often seen in high
temperature superconductors where there is oxygen
disorder\cite{Timusk95,Basov98} which, in Hg-1223, is expected more in
optimally doped samples than in underdoped ones\cite{Sacuto98}.  In agreement
with this, the peak in the optical conductivity spectra of the underdoped
sample, shown in the inset of Fig.\ 2, is narrower and weaker.  It should
also be noted that Hg-1223 crystals are prone to surface deterioration with
exposure to air\cite{Sacuto}.  The underdoped sample was measured soon after
growth, and it is possible that some of the absorption in the optimally doped
sample, measured a year later, could be attributed to the influence of
parasitic phases\cite{Sacuto}.

The overall infrared spectral weight up to 8000~cm$^{-1}$, expressed as a
squared plasma frequency from the conductivity partial sum rule, is
$I(8000)=\int_0^{8000}\sigma_1(\omega)d\omega=4.3\times10^8$~cm$^{-2}$. This
is a factor of 1.3 higher than the comparable figure for the $a$-axis spectral
weight of the two-layer Y-123 at optimal doping and a factor of 2.2 higher
than that for Bi-2212\cite{Puchkov96}. In terms of spectral weight per plane
copper, the Hg-1223 value is higher by factors of 1.33 and 1.53 respectively.

\begin{figure}[t]
\leavevmode
\epsfxsize=\columnwidth
\centerline{\epsffile{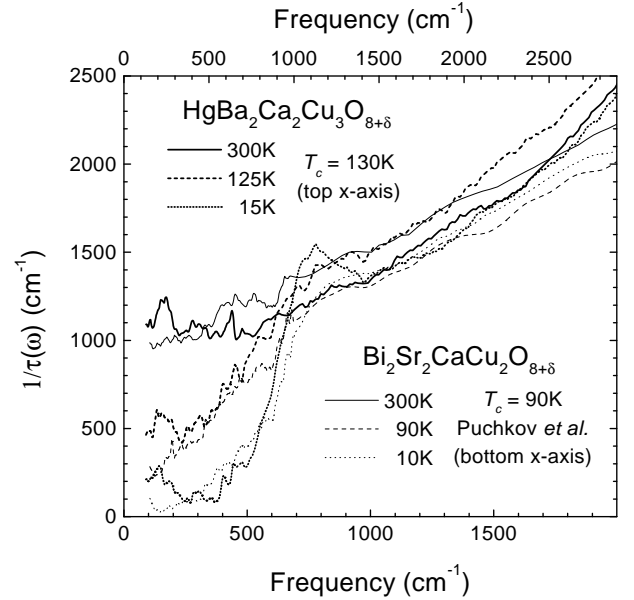}}
\vspace{0.1in}
 \caption{A comparison of the frequency-dependent scattering rate of Hg-1223
with that of Bi-2212 from Puchkov {\it et al.}.  The frequency scales differ
by a factor of $135/93=1.45$, and a frequency-independent scattering rate of
300~cm$^{-1}$ has been subtracted from the Hg-1223 data.}
\label{figure4}
\end{figure}

The frequency-dependent scattering rate was calculated using the extended
Drude model and is shown in Fig.\ 3 for the optimally doped sample. Here the
scattering rate gap is clearly seen as a temperature-dependent depression of
the scattering rate below 1000~cm$^{-1}$.  There is a substantial residual
scattering of 300~cm$^{-1}$ at low frequencies which is consistent with the
300~cm$^{-1}$ width of the low frequency peak in the optical conductivity at
low temperature. The scattering rate of the underdoped sample is shown in the
inset of Fig.\ 3.  It shows no residual scattering, and is about
300~cm$^{-1}$ lower than that of the optimally doped sample over the entire
range shown.

The two samples also show different scattering rate gap onset temperatures.
For the optimally doped sample $125~\mbox{K}<T^*<225$~K, while for the
underdoped sample $T^*>300$~K.  These values are consistent with the dc
resistivity\cite{Carrington94} and NMR\cite{Julien96} values, and the doping
dependence is similar to that of other high temperature
superconductors\cite{Puchkov96}.

The scattering rate gap feature at 1000~cm$^{-1}$ in Hg-1223 is at a
considerably higher frequency than the corresponding features in Y-123,
Y-124, Bi-2212 and Tl2201 which are all at about
700~cm$^{-1}$\cite{Puchkov96}. Since these superconductors also have similar
values of $T_c\approx 93$~K, the scattering rate gaps are in the same ratio
($1000/700=1.43$) as the transition temperatures ($135/93=1.45$).  This is 
consistent with an interpretation by Williams {\it et al.}\cite{Williams97} 
of NMR data on several high temperature superconductors.

Julien {\it et al.} analyzing hyperfine NMR shift data find a much higher
transferred hyperfine coupling constant $B$ in Hg-1223 than in Y-123,
suggesting a higher in-plane superexchange $J$ between the copper
spins\cite{Julien96}. With our finding of a larger scattering rate gap in
the Hg-1223 material these results support the idea of a superconducting
mechanism closely tied to the antiferromagnetism of the copper oxygen planes.

Fig.\ 4 shows a comparison of the scattering rate of Hg-1223 with that of
Bi-2212 from Puchkov {\it et al.}\cite{Puchkov96}. The frequency scales 
differ by the ratio of the values of maximum $T_c$, 135/93,
and a frequency independent scattering rate of 300~cm$^{-1}$ has been
subtracted from the Hg-1223 data to take the residual scattering into account.
The similarity of the scattering rate spectra of the two-layer Bi based
superconductor with those of the three-layer Hg material is striking. A
slight overshoot like the one seen in the 1200~cm$^{-1}$ region for Hg-1223
is also seen in several other optimally doped materials\cite{Puchkov96}.

\begin{figure}[t]
\leavevmode
\epsfxsize=\columnwidth
\centerline{\epsffile{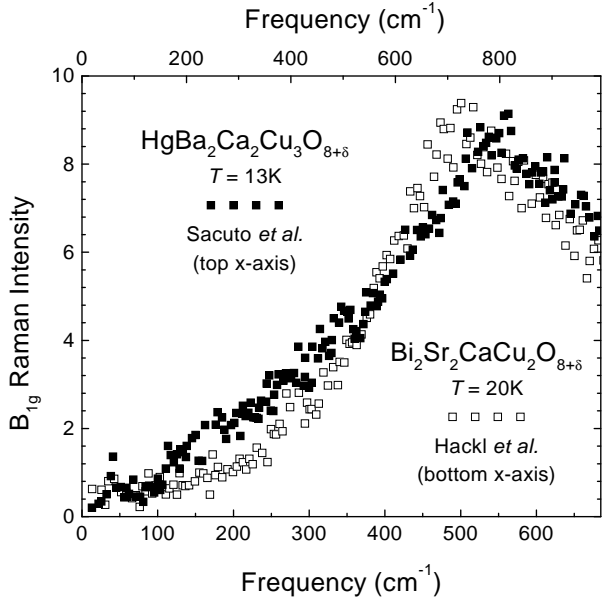}}
\vspace{0.1in}
 \caption{A comparison of the B$_{1g}$ Raman Intensity of Hg-1223 measured by
Sacuto {\it et al.} with that of Bi-2212 measure by Hackl {\it et al.}. The
frequency scales differ by a factor of $135/93=1.45.$}
\label{figure5}
\end{figure}

A similar scaling between the gap and $T_c$ can be seen in B$_{1g}$ Raman
spectra. Fig.\ 5 is a comparison of the B$_{1g}$ Raman intensity of nearly 
optimally doped Hg-1223 measured by Sacuto {\it et al.}\cite{Sacuto98} with 
that of Bi-2212 measured by Hackl {\it et al.}\cite{Hackl96}.  Again the 
frequency scales differ by a factor of $135/93=1.45$.  The positions of the 
maxima at 500 and 800~cm$^{-1}$ are proportional to the superconducting gaps.

There are two low $T_c$ cuprate superconductors for which we have direct 
measurements of the size of the scattering rate gap.  
Bi$_2$Sr$_{2-x}$La$_x$CuO$_{6-\delta}$ (Bi-2201) has a maximum $T_c$ of 29~K. 
The only estimate of gap size for this material comes from an ARPES 
measurement by Harris {\it et al.}\cite{harris97}. They find that the gap 
in Bi-2201 is roughly three times smaller than the gap in Bi-2212 and hence 
does seems to scale with maximum $T_c$.

La$_{2-x}$Sr$_x$CuO$_4$, with a gap of about 700~cm$^{-1}$ and a maximum $T_c$ 
of only 40~K, is the one material that violates the universal scaling between
scattering rate gap and maximum $T_c$. Many of its other properties,
however, are also inconsistent with the gap properties observed in other
cuprate superconductors, and it has been suggested that this may be due to
the presence of paramagnetic centres intrinsic to the random
alloy\cite{Timusk99}. It is reasonable to conclude that 
La$_{2-x}$Sr$_x$CuO$_4$ should be treated as a special case.

A universal scaling between scattering rate gap and maximum $T_c$ is far
from trivial even in the superconducting state. For example, in
conventional electron-phonon mediated superconductors, 
$2\Delta / k_bT_c =3.5$ for the lowest $T_c$ materials, but {\it rises}
with $T_c$ to as high as 5.2\cite{carbotte90}. We see no change in
the gap ratio as maximum $T_c$ is enhanced by 40\%. In order to further
investigate this remarkable behaviour of the cuprate superconductors,
estimates of gap size need to be obtained for more systems with $T_c$
significantly different from 93~K, particularly the two-layer Hg compound,
Hg-1212, and the two- and three-layer Tl compounds, Tl-2212 and Tl-2223.

We are grateful to A.~Sacuto for helpful discussions. The work at McMaster
University was supported by the Natural Sciences and Engineering Research
Council of Canada.

\end{document}